# Topology of Networks in Generalized Musical Spaces

## Marco Buongiorno Nardelli[1,2,3,#]

[1] CEMI – Center for Experimental Music and Intermedia, University of North Texas, Denton, TX, 76203, USA.

[2] iARTA – Initiative for Advanced Research in Technology and the Arts, University of North Texas, Denton, TX, 76203, USA.

[3] Department of Physics, University of North Texas, Denton, TX, 76203, USA.

[#] currently in residence at Fondation IMéRA of Aix-Marseille Université and the CNRS-PRISM Laboratory, Marseille, 13004, France

E-mail: mbn@unt.edu, website: www.materialssoundmusic.com

**Abstract:** The abstraction of musical structures (notes, melodies, chords, harmonic or rhythmic progressions, etc.) as mathematical objects in a geometrical space is one of the great accomplishments of contemporary music theory. Building on this foundation, I generalize the concept of musical spaces as networks and derive functional principles of compositional design by the direct analysis of the network topology. This approach provides a novel framework for the analysis and quantification of similarity of musical objects and structures, and suggests a way to relate such measures to the human perception of different musical entities. Finally, the analysis of a single work or a corpus of compositions as complex networks provides alternative ways of interpreting the compositional process of a composer by quantifying emergent behaviors with well-established statistical mechanics techniques. Interpreting the latter as probabilistic randomness in the network, I develop novel compositional design frameworks that are central to my own artistic research.

**One Sentence Summary:** Network theory is an innovative tool for the classification of generalized musical spaces and provides a framework for the discovery or generation of functional principles of compositional design.

**Introduction.** Network analysis methods exploit the use of graphs or networks as convenient tools for modeling relations in large data sets. If the elements of a data set are thought of as "nodes", then the emergence of pairwise relations between them, "edges", yields a network representation of the underlying set. Similarly to social networks, biological networks and other well-known real-world complex networks, entire dataset of musical structures can be treated as a networks, where each individual musical entity (pitch class set, chord, rhythmic progression, etc.) is represented by a node, and a pair of nodes is connected by a link if the respective two objects exhibit a certain level of similarity according to a specified quantitative metric. Pairwise similarity relations between nodes are thus defined through the introduction of a measure of "distance" in the network: a "metric". [1] As in more well-known social or biological networks, individual nodes are connected if they share a certain property or characteristic (i.e., in a social network people are connected according to their acquaintances, collaborations, common interests, etc.) Clearly, different properties of interest can determine whether a pair of nodes is connected; therefore, different networks connecting the same set of nodes can be generated.

Network representations of musical structures are not new: from the circle of fifths [2], to the Tonnentz [3], and recent works on the spiral array model of pitch space, [4] the geometry of musical chords [5] and generalized voice-leading spaces [6] [7], music theorists, musicians and composers have been investigating how these structures can be combined in explaining the relations between harmony and counterpoint, the foundations of western music. The original contribution of this paper is in the introduction of the representation of musical spaces as large-scale statistical mechanics networks: uncovering their topological structure is a fundamental step to understand their underlying organizing principles, and to unveil how classifications or rule-based frameworks (such as common-practice harmony, for instance) can be interpreted as emerging phenomena in a complex network system.

The paper is organized as follows: I will first illustrate my approach by introducing two different metrics in the pitch class space: one based on the concept of distance between "interval vectors", the other on the distance in voice-leading space. I will then discuss the definition of "rules" in the musical space and introduce the idea of composition as emerging behavior in a complex network. Finally, I will briefly discuss the extension of this method to the classification and analysis of rhythmic progressions. A study on timbre, which requires a representation not only of perceived fundamental frequencies but also of all component partials, is currently ongoing and will be the subject of a future publication.

The computational tools that have been developed for this work are at the core of my personal compositional practice and are collected on the public GitHub repository music♩ntwrk at: https://github.com/marcobn/musicntwrk. An extensive description of the API of this software is available in the Supplemental Information (SI).

**Networks in generalized musical spaces: pitch class sets.** Most of the existing approaches to the geometry of musical spaces focus on harmonic relations among pitches, and attempt to define the interrelations within musical progressions as a collection of transformations obtained by the application of the five operations of octave equivalence (O), permutation (P), transposition (T), inversion (I) and cardinality change (C) to a given pitch class set. [8]

To build a musical network for the pitch space, I start from the *ansatz* that the totality of a musical space can be constructed by an all-combinatorial approach: the super-set of all possible tuples of $N_C$ (cardinality) integers out of $N_P$ total pitches, with $N_C = 1,…,N_P$ – mathematically enumerated by the binomial coefficient. Obviously, these are very large spaces: the traditional chromatic set of the 12-tone equal temperament system (12TET) produces 4,096 possible choices; extending to quarter tones (24TET) we have 16,777,216 combinations, and from the 88 keys of a piano we can produce a staggering $3.1\times10^{26}$ combinations, three orders of magnitude more than the number of units in one mole of any substance, the Avogadro number! These combinations do not allow repetitions of pitches; thus, they form already a drastic geometrical abstraction of the totality of phase-space that is available for music creation. Further abstractions based on a variety of considerations can reduce the dimension of these spaces. In music theory one relies on the five OPTIC transformations to define classes of independent pitch sets [8]*,* [9] and derive classifications that can be used as analytical or compositional tools. By imposing such equivalence classes, one can reduce the 12TET combinatorial space to a mere 238 pitch class sets [10], or the 24TET to 365,588 sets. This approach can equally describe any arbitrary note sets, tuning or temperament systems.

Any set of operations that abstracts the super-set of the musical space defines a "dictionary" of the set space. This is an ordered list of {*label*, *pitch set*, … *other descriptors of the set*} elements that is exhaustive of all the allowed combinatorial possibilities for that space. The elements of the dictionary can then be interpreted as the "nodes" of a deterministic (synthetic) network by defining a proper metric within that space. "*label*" here represents the generalization of the Forte classification scheme [10] for arbitrary $N_P$, by sorting the combinatorial sets in ascending order, and eventually identifying distinct sets that share one or more identical descriptors, as in Z-related sets (pcs with same interval vector but different prime form).

To navigate this network I first introduce a metric based on the Euclidean distance (a generalized multidimensional Pythagoras theorem) between vectors of integers: here I choose the interval vectors as nodes of the network, the array of natural numbers which summarize the intervals present in a pitch class set, one of its fundamental descriptors (other descriptors can be defined along similar lines [11] but will not be considered here). See SI (eq. S1) for a formal definition of this distance operator.

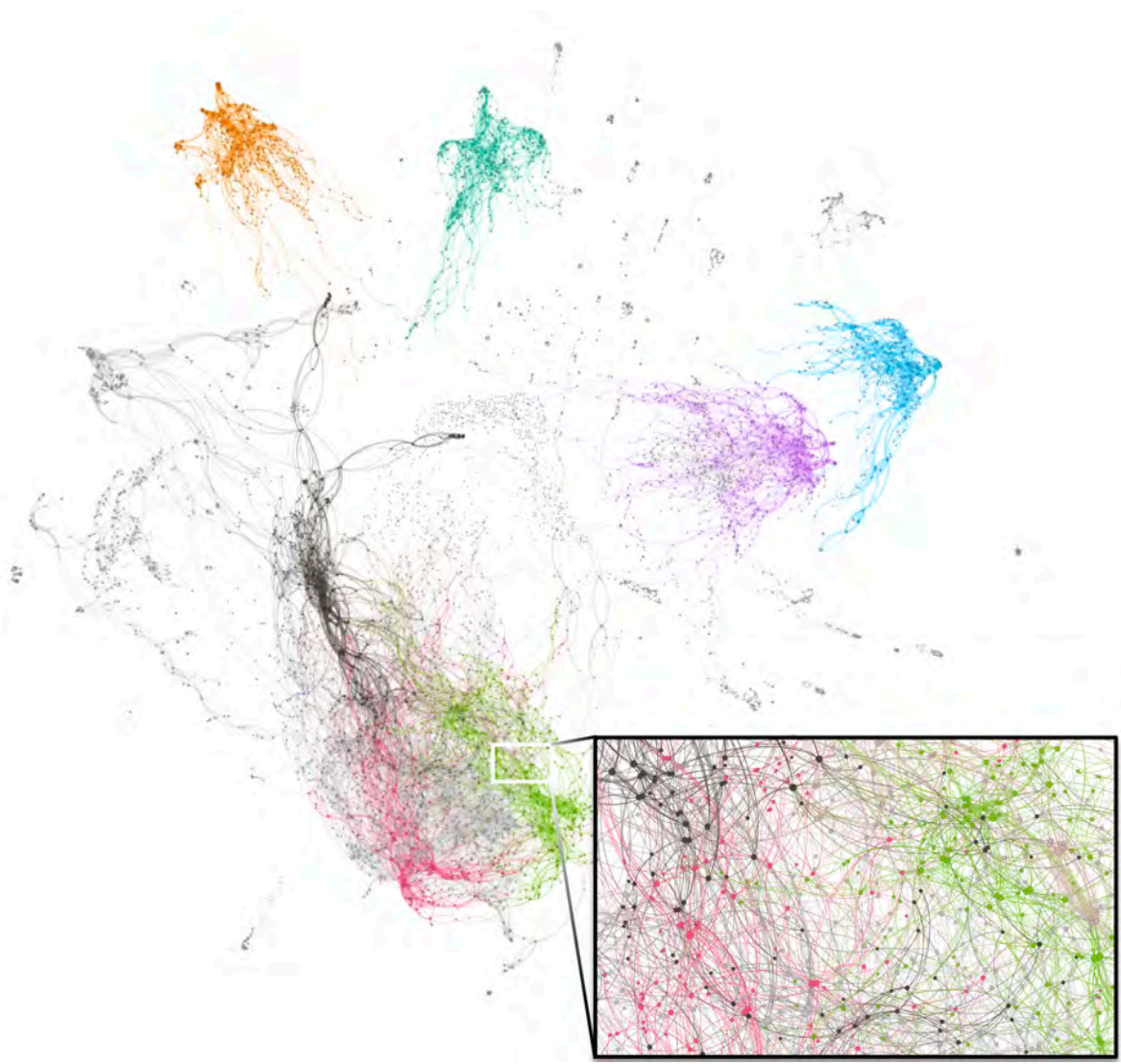

**Fig. 1.** Interval vector-based network of all the 7 note scales that can be constructed in the 24TET super-set for distances $d=\sqrt{2}$ (nearest neighbors only). Nodes are grouped by the degree of connections (number of connections per node) and color-coded according to their modularity class. The inset shows the high degree of interconnection even for a single distance threshold.

Given the integer character of the underlying vectors, distances are "quantized": only discrete values are allowed. This observation suggests the introduction of a new class of operators, $\mathbf{O}(\{n_i\})$, see SI eq. S3, that raise or lower by an integer $n$ the $i^{th}$ component of an interval vector. These operators will play a major role in the network analysis that follows.

As an illustration of this technique, I show in Fig. 1 the network of all the 7 note scales that can be constructed in the 24TET super-set: 7478 unique nodes (prime forms – OPTI equivalence classes [10] [12]) with 1715 Z-related sets. I map this network for distances $d=\sqrt{2}$, or equivalently, by applying the $\mathbf{O}(1,1)$ operator, which raises or lowers two elements of the interval vector by 1.

Although the above classification could have been obtained also by other mathematical means, the novelty here is that having a network representation of a musical space allows us to apply well-established techniques of statistical mechanics for the analysis of large-scale networks and to quantitatively examine the structure of relationships between pitch classes. Indeed, given a network we can perform many statistical operations that shed light on the internal structure of the data. In this work I will consider only two of such measures, degree centrality and modularity class. The degree of a node is measured by the number of edges that depart from it. It is a local measure of the relative "importance" of a node in the network. Modularity is a measure of the strength of division of a network into communities: high modularity (above 0.6 in a scale from 0 to 1) corresponds to networks that have a clearly visible community structure. [13] In the case of the network of Fig. 1, I measure an average degree of 5.33 and a modularity of 0.865, clearly manifest in the high degree of separation between regions of different colors. Isolating communities provides a way to operate within regions of higher similarity, and thus, in this particular case, of closer harmonic content.

**Networks in generalized musical spaces: voice leading.** Let's now move to the construction of networks based on the voice-leading distance measure (see SI and eq. S2 for the definition of this particular metric). Here the nodes of the network represent individual pcs (chords) and their relationships (edges) are defined by their distance in the harmonic space. For illustrative purposes, I will start by restricting our analysis to the space of all the triads in 12TET. Let's first consider the super-set of octave-equivalent normal forms (prime forms plus inversions – OPT equivalence classes). In this space, a major and a minor chord are considered different although they can be reduced to the same prime form. This is the space that best abstracts three-part counterpoint in common practice harmony. The network restricted to nearest neighbors ($\mathbf{O}(1)$) is isomorphic to the orbifold of the quotient space $\mathbb{T}^2/\mathcal{S}_3$ as derived in Ref. [6] and shown in Fig. S3. This result demonstrates clearly the power of our approach: our network analysis is confirming relations between sets when these are known, a proof of its reliability, and at the same time provides a direct method to explore musical spaces beyond the simplest abstractions.

To illustrate this argument even further, let's release the constraint on the transposition equivalence class and derive the network of the panchromatic three-parts voice-leading space in 12TET. The four panels in Fig. 2 show networks sliced at different distance thresholds (data to reproduce every network are available as SI). Network a) is the orbifold of minimal distance neighbors. In a common practice harmony framework, it is the extension of the $\mathbb{T}^2/\mathcal{S}_3$ orbifold to every major and minor key. Network a) displays an average node degree of 4.9 with a modularity index of 0.56. b) is the network with edges at $d(\mathbf{x},\mathbf{y})=\sqrt{2}$ corresponding to the operator $\mathbf{O}(1,1)$. Interestingly, for operators $\mathbf{O}(\{n_i\})$ of higher order, the topology of the network can be drastically altered. In this case, the network is split into two disconnected orbifolds that have as centers the augmented C,D and C#,Eb triads, respectively, with an average degree of 8.8 and a modularity of 0.57.

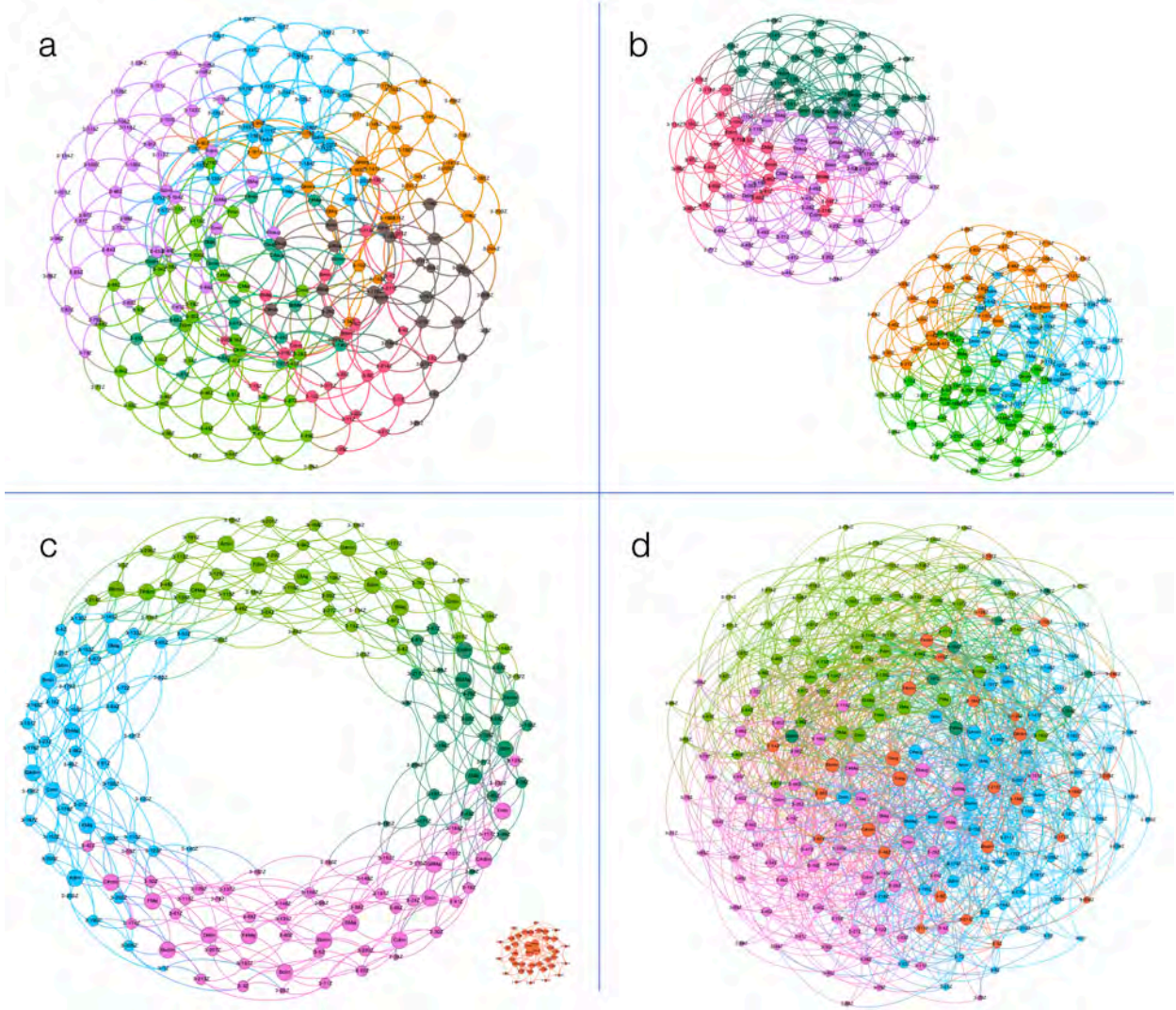

**Fig. 2.** Networks of the panchromatic three-parts voice-leading space in 12TET. a) $d(\mathbf{x},\mathbf{y})=1$ - $\mathbf{O}(1)$; b) $d(\mathbf{x},\mathbf{y})=\sqrt{2}$ - $\mathbf{O}(1,1)$; c) $d(\mathbf{x},\mathbf{y})=\sqrt{3}$ - $\mathbf{O}(1,1,1)$; d) $d(\mathbf{x},\mathbf{y})=\sqrt{5}$ - $\mathbf{O}(1,2)$.

Similarly, in c) the network, defined by $d(\mathbf{x},\mathbf{y})=\sqrt{3}$ - $\mathbf{O}(1,1,1)$, is split into a large torus that excludes all augmented triads and their close relatives (degree = 5.9, modularity = 0.63; finally, in d) the topology of the orbifold is recovered by $d(\mathbf{x},\mathbf{y})=2$ - $\mathbf{O}(2)$, with degree = 13.94 and modularity = 0.26. The structure of the network built under any of the $\mathbf{O}(\{n_i\})$ operators reflects particular classes of functional properties of chord progressions. Some of the structures outlined in networks a-d are summarized in Table 1. One must note that the $\mathbf{O}(\{n_i\})$ operators contain and generalize the operators of the neo-Riemannian triadic theory: **P,L**=$\mathbf{O}(1)$, **R**=$\mathbf{O}(2)$, **N,S**=$\mathbf{O}(1,1)$, **H**=$\mathbf{O}(1,1,1)$. [14] More generally, any progression in a given musical space can be obtained by the successive application of $\mathbf{O}(\{n_i\})$ operators, thus creating the desired sequence.

| Operator | Distance | Example of functional chord progression |
|---|---|---|
| **O**(1) | 1 | aug ⇒ Maj ⇒ min ⇒ dim |
| **O**(2) | 2 | Maj to 7 progression (*i.e.* C ⇒ C7) |
| **O**(1,2) | $\sqrt{5}$ | IV-I progression |
| **O**(1,1,2) | $\sqrt{6}$ | V7-I progression |
| **O**(1,2,2) | 3 | I-ii, vii°-I and IV-V progression |

**Table 1.** Example of the functional chord progressions from selected distance operators.

Finally, the distance operators find a direct application in the definition of "parsimony", as the average of the weight (inverse of the distance between two chords) along the progression provides a measure of the "motion" of the sequence. See SI for a more detailed discussion.

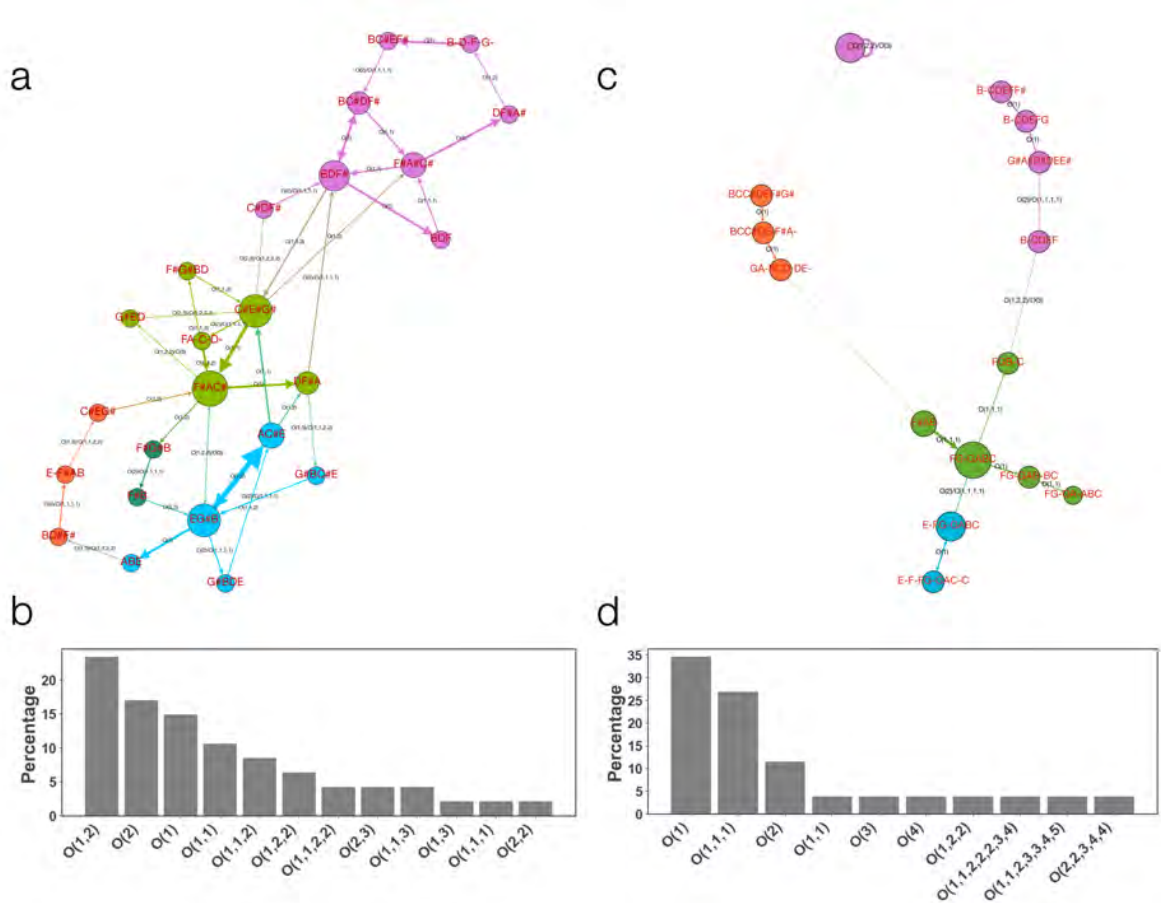

**Fig. 3.** a) voice leading network of J.S. Bach's chorale from the Cantata "Erfreut euch, ihr Herzen", BWV 66 (1724); b) probability distribution of occurrences of specific distance operators in the chorale; c) voice leading network of the sixth movement (Sehr langsam) of A. Schoenberg "Sechs kleine Klavierstücke", Op. 19 (1913); d) probability distribution of occurrences of specific distance operators in Schoenberg's piece. A video animation of the evolution of both progressions is available in SM.

**"Composition" as emergent behavior in a complex network.** All the networks in Figs. 1 and 2 are "synthetic": they are the result of the deterministic application of equivalence classes or other discrete operators to an all-combinatorial super-set. As such they provide us with a complete account of the structure and topology of arbitrary musical spaces but do not add any additional understanding beyond their geometrical structure. In the process of music making, the composer necessarily builds her own harmonic and melodic sequence by making specific choices on the underlying structure of the network based on a wide variety of considerations: from esthetic to functional, personal or programmatical. Moreover, the network is generally already sliced according to generic rules of harmonic progression, where, for instance, only selected edges are allowed (i.e. common practice harmony).

In doing this, the deterministic fabric of the absolute geometrical space is shredded by probabilistic choices: the network becomes complex. This is the process at the core of the creation of any musical work: the composer, by introducing probabilistic choices in the creation of her own version of the network, induces emergent behaviors that manifest themselves in the harmonic and contrapuntal framework of the piece.

All these concepts can be observed directly in the analysis of a score: as an illustration we show in Fig. 3 the network representations of two works of contrasting characteristics: J.S. Bach's chorale from the Cantata "Erfreut euch, ihr Herzen", BWV 66 and the sixth movement (sehr langsam) of A. Schoenberg "Sechs kleine Klavierstücke", Op. 19. The edges here are directional, to indicate the pcs progression in the piece. A simple statistical analysis shows that Bach's chorale network, Fig. 3a, has an average degree of 1.70 per node, and a modularity index of 0.46: the distribution of edges is very sparse, and many nodes are visited numerous times. The modularity index clearly individuates the broad tonal areas visited in the short piece (color-coded in Fig 3a): we start from A Major, modulate to F# minor, then back to A Major with a pass-through C# minor, then C# Major and finally ending in F# Major (see movie in SI). Analysis of the distribution of the $\mathbf{O}(\{n_i\})$ operators (Fig. 3b) highlights the predominance of $\mathbf{O}(1,2)$ and $\mathbf{O}(2)$, $\mathbf{O}(1)$, $\mathbf{O}(1,1,2)$ and $\mathbf{O}(1,2,2)$ in the voice leading space (see Table 1). To further generalize this analysis, I created a voice-leading network of the full corpus of the 371 Bach's chorales in the `music21` [15] database: the results are shown in Fig. 4 and confirm what was observed in BWV 66: the five operators in Table 1 constitute almost 80% of all voice leadings in the corpus – the core operators of common practice harmony as emergent behaviors in a complex network!

In contrast, Schoenberg's network displays a strikingly different topology (Fig. 3c): the edges form a loop and each node has much fewer connections (average degree of 1.27). The progressions are largely chromatic, with $\mathbf{O}(1)$ and $\mathbf{O}(1,1,1)$ used extensively (Fig. 3d) and coexists with leaps of large distances (7, $\sqrt{65}$ and $\sqrt{39}$) in the voice leadings, corresponding to movements mostly from single pitches to large chords, an evidence of the broad range of cardinalities in the pcs (between 1 and 8 vs. the largely triadic harmony of Bach's chorale). Clearly, new relations that replace the framework of classical harmony emerge from this network. However, a modularity of 0.48 demonstrate the resilience of a compositional design that is very classical: the work is still centered strongly on a harmonic center (the incomplete dominant 7[th] chord F#AB – Forte class 3-7) and uses the FF#GABC cluster (Forte class 6-Z12) as a pivot point of most progressions (see movie in SI).

**Networks in generalized musical spaces: rhythmic progressions.** Before concluding, I would like to briefly discuss the extension of this approach to networks of rhythmic progressions. A great amount of research has been devoted to this subject and rather than providing a comprehensive discussion on past accomplishments, (see for instance Ref. [16]) I will focus on how the network approach can be applied in this framework. We need to develop dictionaries of rhythmic objects that can constitute the nodes of an eventual network, and a metric, to define their relative distance in an appropriate rhythmic space. I use two different representations of rhythmic space: the first is based on sequences of $N_P$ note durations where I look for all possible combinations of arbitrary cardinality, the rhythmic cells; the second is instead based on a fixed length (for instance, a 4/4 measure) and looks for all possible sequences that can be constructed to fill that particular length. Within these spaces of rhythmic sequences, I define two descriptors. One is an extension of the concept of pitch class set interval vector, i.e the total relative duration ratio content of the sequence (how many intervals of a quarter (or eight, sixteen, etc.) note can be constructed with the note durations of that cell). I call this the "duration vector". The second is the inter-onset duration interval content of the rhythmic cell, [17] the sequence of the duration intervals between the onsets of the individual notes. With these descriptors we can then introduce the measure of distance in the same fashion as for pcs networks: the Euclidean distance between the duration or interval vectors (see SI and discussion therein). Although, so far, I have used this approach to rhythmic networks only as a compositional tool, the reader is encouraged to familiarize with the computational tools in the music∮ntwrk repository and apply the method for the analysis of rhythmic structures.

**Conclusions.** The analytical process outlined above is first and foremost a methodology that I have developed for my own compositional practice. It can be clearly generalized for the generation of algorithmically-based music composition agents: current generators for complex networks that include, among others, the well know models of Erdös-Rényi (probability of edge creation) [18] or Barabási-Albert (probabilistic distribution of the number of edges to attach from a new node) [1], can be used for the generation of such complex networks. For instance, by using the Barabási-Albert algorithm of preferential attachment with a probability distribution of my choice, I can generate a sliced network that would correspond to totally new harmonic progression rules and, from there, generate innumerable variations on similar structures. This approach provides internal coherence within the framework and, eventually, could lead to novel pathways for the design of algorithmic agents of automatic composition (see Ref. [19] for a comprehensive review of algorithmic composition and artificial intelligence). These procedures are becoming essential in my artistic research. As an illustration, the reader can find the score and the compositional notebook of "Le Reseau de Ton Souvenir" a suite for solo Alto Recorder that uses, at its foundation, harmonic hierarchies and rhythmic sequences based on the complex network analysis outlined in this paper.

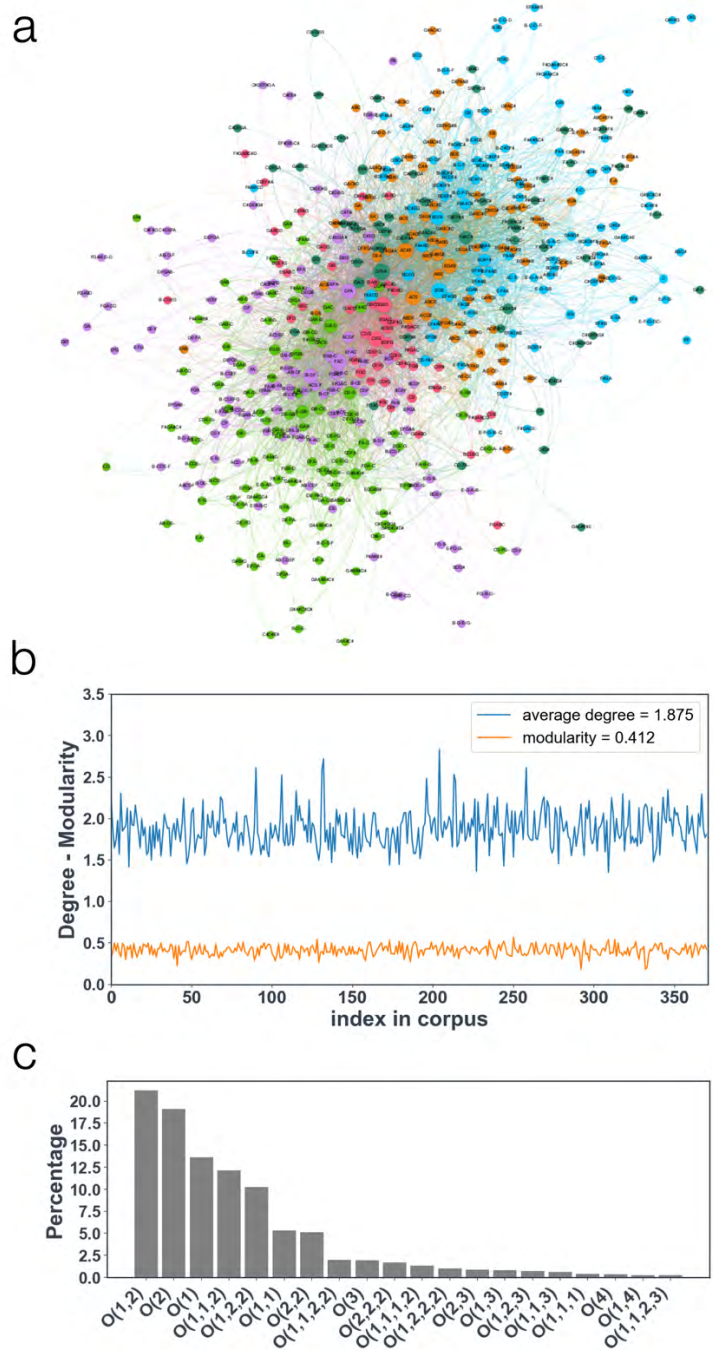

**Fig. 4.**
a) Voice leading network of the full corpus of J.S. Bach chorales; b) average degree and modularity class across the corpus; c) probability distribution of occurrences of specific distance operators in the corpus. The average degree (1.88) and modularity (0.41) calculated for each individual chorale (b), demonstrate a high degree of homogeneity in the voice-leading structures across the corpus; the full network in (a) manifests the centrality of preferential keys.

**Acknowledgments:** I acknowledge useful discussions with Jon Nelson, Joseph Klein, Scot Gresham-Lancaster, David Bard-Schwarz, Roger Malina and Alexander Veremyer. I am grateful to the Fondation IMéRA of Aix-Marseille Université and the CNRS-PRISM Laboratory for the continuous support and encouragement during the writing of this paper. **Competing interests:** the author declares no competing interests; **Data and materials availability:** All data is available in the main text or as supplementary material. The python modules for the generation and analysis of networks in generalized musical spaces, `musicntwrk`, is freely available at https://github.com/marcobn/musicntwrk and its API is described in SI.

**Supplementary Materials:**

Metrics in generalized musical spaces
Parsimony as measure of distance (Fig. S1)
Methods: the `musicntwrk` API
Fig. S2
Captions for Movie S1
Captions for Data S1 to S8
Movie S1
Data S1 to S8
Score of "Le Reseau de Ton Souvenir" for Alto Recorder

**Biographical information**

Marco Buongiorno Nardelli is University Distinguished Research Professor at the University of North Texas: composer, media artist, flutist, and computational materials physicist, he is a member of CEMI, the Center for Experimental Music and Intermedia, and iARTA, the Initiative for Advanced Research in Technology and the Arts. He is a Fellow of the American Physical Society and of the Institute of Physics, and a Parma Recordings artist. From February to July 2019 he is at Fondation IMéRA (Institute Méditerranéen d'Etudes Avancées) of Aix-Marseille Université and at the PRISM (Perception, Representations, Image, Sound, Music) Laboratory of the CNRS as the first holder of the PRISM/IMéRA Chaire.

# Supplementary Materials for

## Topology of Networks in Generalized Musical Spaces

**Marco Buongiorno Nardelli**[1,2,3,#]

[1] CEMI – Center for Experimental Music and Intermedia, University of North Texas, Denton, TX, 76203, USA.

[2] iARTA – Initiative for Advanced Research in Technology and the Arts, University of North Texas, Denton, TX, 76203, USA.

[3] Department of Physics, University of North Texas, Denton, TX, 76203, USA.

[#] currently in residence at Fondation IMéRA of Aix-Marseille Université and the CNRS-PRISM Laboratory, Marseille, 13004, France

E-mail: mbn@unt.edu, website: www.materialssoundmusic.com

**This PDF file includes:**

    Metrics in generalized musical spaces
    Parsimony as measure of distance (Fig. S1)
    Methods: the `musicntwrk` API
    Fig. S2
    Captions for Movie S1
    Captions for Data S1 to S8

**Other Supplementary Materials for this manuscript include the following:**

    Movie S1
    Data S1 to S8
    Score of "Le Reseau de Ton Souvenir" for Alto Recorder

**Metrics in generalized musical spaces**

A. <u>Interval vectors</u>: The distance operator for interval vectors can be written as:

$$d(\mathbf{x},\mathbf{y}) = \sqrt{\sum_i (\mathbf{x}_i - \mathbf{y}_i)^2}, \tag{S1}$$

where **x** and **y** are interval vectors of dimension INT($N_P$/2), where $N_P$ is the dimension of the PCS (number of total pitches). This is a quantity that measures the change in the harmonic content of two pcs, and thus contains a quantification of the rules of harmony in arbitrary musical spaces.

B. <u>Voice leading</u>: the distance operator for pcs provides instead a quantification of voice leading, the study of the linear progression of individual melodic lines at the foundation of counterpoint. For this we use minimal Euclidean voice leading distance [1] for arbitrary TET-notes temperaments (12, 24, etc.):

$$d_{\min}(\mathbf{x},\mathbf{y}) = \min_{\mathbf{TET}_j} \sqrt{\sum_i \left(\vec{x}_i - (\vec{y}_i \pm \mathbf{TET}_j)\right)^2}. \tag{S2}$$

Here $\vec{x}_i$ and $\vec{y}_i$ are the pcs in ascending order and $\mathbf{TET}_j = (0,0,...,0,\pm \text{TET}, 0, ...)$, a vector of dimension $N_C$ that raises or lowers the $j^{th}$ pitch of the ordered pcs by TET. It is easy to verify that this definition of distance operator is equivalent to finding the minimal distance between all possible cyclic permutations of the pcs. Such definition is easily extended to non-bijective voice-leadings.

C. <u>Distance operators</u>: $\mathbf{O}(\{n_i\})$ - they raise or lower by an integer $n$ the $i^{th}$ component of a vector. In the interval vector space, these are vector operators of dimension INT($N_P$/2); in the voice-leading space, they have dimension $N_C$. With this definition, if $x$ is transformed into $y$ by $\mathbf{O}(\{n_i\})$, then:

$$d(\mathbf{x},\mathbf{y}) = \sqrt{\sum_i n_i^2}. \tag{S3}$$

In the current implementation of the method, the distance operator is assumed to be raising or lowering by the specified amount one of the pitches of the pcs with no information on the positional ordering (*i* is left unspecified). For instance, $\mathbf{O}(\mathbf{1})$ applied to the [0,4,7] pcs (C Maj chord) generates the following chords: [0, 3, 7] (C min), [0, 4, 6] (C incomplete half-diminished seventh), [0, 4, 8] (C augmented), [5, 7, 0] (F quartal trichord), [1, 4, 7] (C# diminished), [4, 7, 11] (E min). It is a simple extension to choose a positional specification (specify *i*), in order to obtain a unique (bijective) mapping.

D. <u>Rhythm duration and interval vectors</u>: as explained in the main text, the duration and interval vectors are the two descriptors I have introduced to construct dictionaries of rhythmic sequences. The total relative duration ratio content of the sequence (how many intervals of a quarter (or eight, sixteen, etc.) note can be constructed with the note durations of that cell). I call this the "duration vector". The second is the "interval vector", the inter-onset duration interval content of the rhythmic cell, the sequence of the duration intervals between the onsets of the individual notes. To illustrate the two definitions, let's take the following rhythmic sequence: [1/2, 3/8, 3/8, 1/2, 1/4]: the components of the duration vector will contain 6 intervals of 1/8 and 2 intervals of 1/4; the components of the interval vector are: 1 x 1/4, 2 x 3/8, 2 x 1/2, 3 x 3/4, and 2 x 7/8. Given this definition, a sequence of reference must be always defined in advance. If the sequence of reference in this case is [1/8 1/4 3/8 1/2 5/8 3/4 7/8 1/1], the duration vector is [6, 2, 0, 0, 0, 0, 0, 0] and the interval vector is [0, 1, 2, 2, 0, 3, 2, 0].

As an illustration, we show in Fig. S1, the network of interval vectors built from all possible sequences that can be constructed to fill a rhythmic length of 16/8 (2 measures of 4/4) in groups of cardinality 5 (only 5 notes of complementary duration). Node labels follow a Forte-like classification scheme, where the suffix Z indicates that the rhythmic sequence shares the same interval vector with other sequences, and N indicates if the sequence is non-retrogradable.

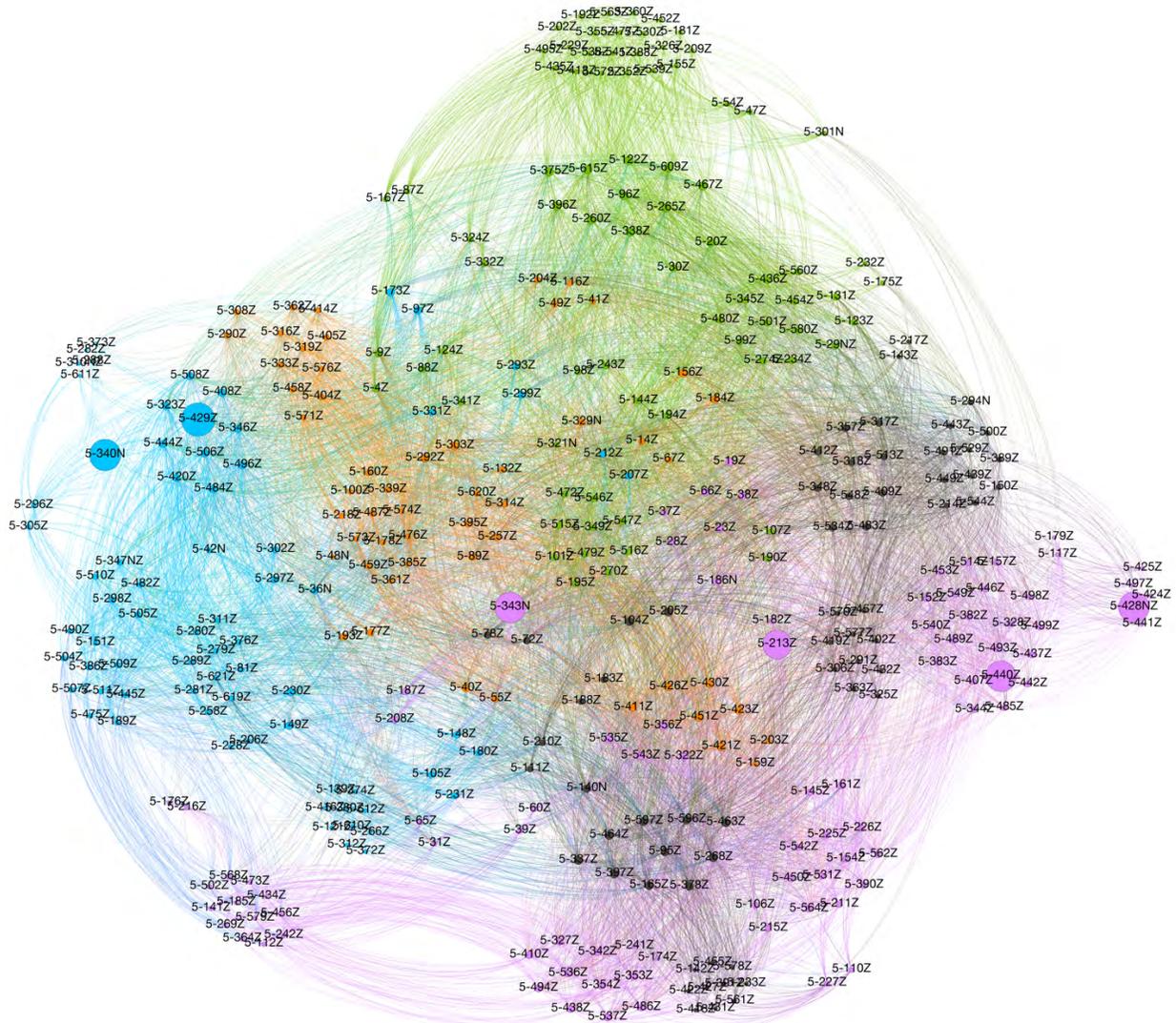

**Fig. S1.** The network of interval vectors built from all possible sequences that can be constructed to fill a rhythmic length of 16/8 (2 measures of 4/4) in groups of cardinality 5. This network contains the six prototypical clave rhythms: shiko, son, soukous, rumba, bossa, and gahu.

**Parsimony as measure of distance**

Since each $\mathbf{O}(\{n_i\})$ is uniquely associated to a distance, we can assign to each step of a harmonic sequence a weight, that for simplicity is defined as the inverse of the distance. The average of the weight along the progression provides a measure of the "motion" of the sequence.

We argue that such measure offers a quantitative estimate of the parsimony (P) of a given voice leading [2] [3] and that can provide a basis for the compilation of a hierarchy of musical entities and their relation to human perception. [4]

As an illustration of this idea, I consider the pcs subspace comprised of all the diatonic triads in the keys of C and G major (Fig. S2a) for distances up to $\sqrt{6}$. I chose four different progressions, labelled A through D and I calculate P: sequence A (G⇒C) is the less parsimonious, since all the harmonic motion happens in one single step ($\mathbf{O}(1,2)$); distributing the motion among passing chords of closer distance, as in B (G⇒Bmin⇒C), C (G⇒Emin⇒C) and D (G⇒Bmin⇒Emin⇒C), increases P, a manifestation of the minimization of the motion from one step to the next (Fig. S2b).

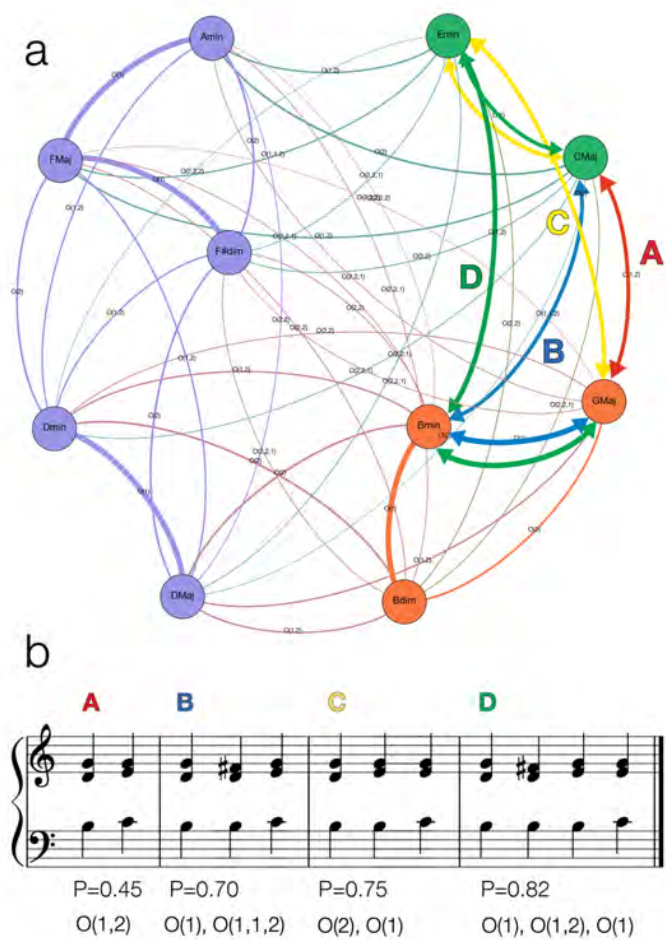

**Fig. S2.** a) pcs subspace comprised of all the diatonic triads in the keys of C and G major for distances up to $\sqrt{6}$, the thickness of the edge lines is proportional to their weight; b) musical illustration of the paths A,B,C and D highlighted in panel a), parsimony values and corresponding operators.

**Methods**

The musicntwrk python modules

All the results discussed in this report have been obtained by the musicntwrk python module, written by the author and available at https://github.com/marcobn/musicntwrk. In the following we briefly describe the software API. musicntwrk is a software written for python 3 and comprises two modules, pcsPy and rhythmPy. It requires installation of the following modules via the "pip install" command:
1. System modules: sys, re, time, os
2. Math modules: numpy, itertools, fractions, gcd, functools
3. Data modules: pandas, sklearn.metrics, networkx, community
4. Music modules: music21
5. Parallel processing: mpi4py

pcsPy is comprised of the PCSet class and its methods (listed below) and a series of functions for pcs network manipulations:

```
class PCSet:
    def __init__(self,pcs,TET=12,UNI=True,ORD=True):
    •   pcs (int)- pitch class set as list or numpy array
    •   TET (int)- number of allowed pitches in the totality of the musical space
        (temperament). Default = 12 tones equal temperament
    •   UNI (logical) - if True, eliminate duplicate pitches (default)
    •   ORD (logical) - if True, sorts the pcs in ascending order (default)
    def normalOrder(self):
    •   Order the pcs according to the most compact ascending scale in pitch-class
        space that spans less than an octave by cycling permutations.
    def normal0Order(self):
    •   As normal order, transposed so that the first pitch is 0
    def transpose(self,t=0):
    •   Transposition by t (int) units (modulo TET)
    def zeroOrder(self):
    •   transposed so that the first pitch is 0
    def inverse(self):
    •   inverse operation: (-pcs modulo TET)
    def primeForm(self):
    •   most compact normal 0 order between pcs and its inverse
    def intervalVector(self):
    •    total interval content of the pcs
    def LISVector(self):
    •   Linear Interval Sequence Vector: sequence of intervals in an ordered pcs
    def operator(self,name):
    •   operate on the pcs with a distance operator
            •   name (str) - name of the operator O({n_i})
    def forteClass(self):
    •   Name of pcs according to the Forte classification scheme (only for TET=12)
    def jazzChord(self):
    •   Name of pcs as chord in a jazz chart (only for TET=12 and cardinalities ≤ 7)
    def commonName(self):
    •   Display common name of pcs (music21 function - only for TET=12)
    def commonNamePrime(self):
    •   As above, for prime forms
```

```python
def nameWithPitchOrd(self):
```
- Name of chord with first pitch of pcs in normal order
```python
def nameWithPitch(self):
```
- Name of chord with first pitch of pcs
```python
def displayNotes(self,xml=False,prime=False):
```
- Display pcs in score in musicxml format. If prime is True, display the prime form.

Network functions:

```python
def pcsDictionary(Nc,order=0,TET=12,row=False,a=np.array(None)):
```
- Generate the dictionary of all possible pcs of a given cardinality in a generalized musical space of TET pitches
    - Nc (int)– cardinality
    - order (logical)– if 0 returns pcs in prime form, if 1 retrns pcs in normal order, if 2, returns pcs in normal 0 order
    - row (logical)– if True build dictionary from tone row, if False, build dictionary from all combinatorial pcs of Nc cardinality given the totality of TET.
    - if row = True, a is the list of pitches in the tone row (int)
    - returns the dictionary as pandas DataFrame and the list of all Z-related pcs
```python
def pcsNetwork(input_csv,thup=1.5,thdw=0.0,TET=12,distance='euclidean',col=2,prob=1):
```
- generate the network of pcs based on distances between interval vectors
    - input_csv (str)– file containing the dictionary generated by pcsNetwork
    - thup, thdw (float)– upper and lower thresholds for edge creation
    - distance (str)– choice of norm in the musical space, default is 'euclidean'
    - col = 2 – metric based on interval vector, col = 1 can be used for voice leading networks in spaces of fixed cardinality – NOT RECOMMENDED
    - prob (float)– if ≠ 1, defines the probability of acceptance of any given edge
    - in output it writes the nodes.csv and edges.csv as separate files in csv format
```python
def pcsEgoNetwork(label,input_csv,thup_e=5.0,thdw_e=0.1,thup=1.5,thdw=0.1,TET=12,distance='euclidean'):
```
- network generated from a focal node (**ego**) and the nodes to whom **ego** is directly connected to (**alters**)
    - label (str)– label of the ego node
    - thup_e, thdw_e (float) - upper and lower thresholds for edge creation from ego node
    - thup, thdw (float)– upper and lower thresholds for edge creation among alters
        - in output it writes the nodes_ego.csv, edges_ego.csv and edges_alters.csv as separate files in csv format
```python
def vLeadNetwork(input_csv,thup=1.5,thdw=0.1,TET=12,w=True,distance='euclidean',prob=1):
```
- generation of the network of all minimal voice leadings in a generalized musical space of TET pitches – based on the minimal distance operators – select by distance
    - input_csv (str)– file containing the dictionary generated by pcsNetwork
    - thup, thdw (float)– upper and lower thresholds for edge creation
    - w (logical) – if True it writes the nodes.csv and edges.csv files in csv format
    - returns nodes and edges tables as pandas DataFrames
    - available also in vector form for computational efficiency as vLeadNetworkVec
```python
def vLeadNetworkByName(input_csv,name,TET=12,w=True,distance='euclidean',prob=1):
```

- generation of the network of all minimal voice leadings in a generalized musical space of TET pitches – based on the minimal distance operators – select by name
  - input_csv (str)- file containing the dictionary generated by pcsNetwork
  - name (str) – name of operator for edge creation
  - w (logical) – if True it writes the nodes.csv and edges.csv files in csv format
  - returns nodes and edges tables as pandas DataFrames
  - available also in vector form for computational efficiency as vLeadNetworkByNameVec

```python
def scoreNetwork(seq,TET=12):
```
- generates the directional network of chord progressions from any score in musicxml format
  - seq (int) – list of pcs for each chords extracted from the score

```python
def scoreDictionary(seq,TET=12):
```
- build the dictionary of pcs in any score in musicxml format

```python
def readScore(inputxml,TET=12,music21=False):
```
- read musicxml score and returns chord sequence
  - inputxml (str) – score file
  - music21 (logical) – if True search the music21 corpus

```python
def extractByString(name,label,string):
```
- extract rows of any dictionary (from csv file or pandas DataFrame) according to a particular string in column 'label'
  - name (str or pandas DataFrame) – name of the dictionary
  - string (str) – string to find in column
  - label (str) – name of column of string

```python
def minimalDistance(a,b,TET=12,distance='euclidean'):
```
- calculates the minimal distance between two pcs of same cardinality (bijective)
  - a,b (int) – pcs as lists or numpy arrays

```python
def minimalNoBijDistance(a,b, TET=12,distance='euclidean'):
```
- calculates the minimal distance between two pcs of different cardinality (non bijective) – uses minimalDistance()
  - a,b (int) – pcs as lists or numpy arrays

```python
def opsDictionary(distance):
```
- dictionary of names of distance operators

```python
def opsName (a,b,TET=12):
```
- given two vectors returns the name of the operator that connects them
  - a,b (int) – pcs as lists or numpy arrays
  - returns name as str

```python
def opsCheckByName (a,b,TET=12):
```
- given two vectors returns check if the connecting operator is the one sought for
  - a,b (int) – pcs as lists or numpy arrays
  - returns logical
  - available also in vector form for computational efficiency as opsCheckByNameVec

```python
def opsDistance (name):
```
- returns distance for given operator
  - name (str) – name of operator
  - returns name (str), distance (float)

```python
def Remove(duplicate):
```
- function to remove duplicates from a list

rhythmPy is comprised of the `RHYTHMSeq` class and its methods (listed below) and a series of functions for rhythmic network manipulations:

```
class RHYTHMSeq:
    def __init__( self,rseq,REF='e',ORD=False)
```
- rseq (str/fractions/floats) – rhythm sequence as list of strings/fractions/floats name
- REF (str) - reference duration for prime form – the RHYTHMSeq class contains a dictionary of common duration notes that uses the fraction module for the definitions (implies "import fraction as fr"):
  {'w':fr.Fraction(1,1),'h':fr.Fraction(1,2),'q':fr.Fraction(1,4),\
   'e':fr.Fraction(1,8),'s':fr.Fraction(1/16),'t':fr.Fraction(1,32),\
   'wd':fr.Fraction(3,2),'hd':fr.Fraction(3,4),'qd':fr.Fraction(3,8),\
   'ed':fr.Fraction(3,16),'sd':fr.Fraction(3,32),'qt':fr.Fraction(1,6),\
   'et':fr.Fraction(1,12),'st':fr.Fraction(1,24), 'qq':fr.Fraction(1,5),\
   'eq':fr.Fraction(1,10),'sq':fr.Fraction(1,20)}. This dictionary can be extended by the user on a case by case need.

```
def normalOrder(self)
```
- Order the rhythmic sequence according to the most compact ascending form.

```
def augment(self,t='e')
```
- Augmentation by t units
  - T (str) – duration of augmentation

```
def diminish(self,t='e')
```
- Diminution by t units
  - T (str) – duration of diminution

```
def retrograde(self)
```
- Retrograde operation

```
def isNonRetro(self)
```
- Check if the sequence is not retrogradable

```
def reduce2GCD(self)
```
- reduce the series of fractions to Greatest Common Divisor

```
def primeForm(self)
```
- reduce the series of fractions to prime form

```
def durationVector(self,lseq=None)
```
- total relative duration ratios content of the sequence
  - lseq (list of fractions) – reference list of duration for evaluating interval content – the default list is:
    [fr.Fraction(1/8),fr.Fraction(2/8),fr.Fraction(3/8),\
     fr.Fraction(4/8),fr.Fraction(5/8),fr.Fraction(6/8),\
     fr.Fraction(7/8),fr.Fraction(8/8),fr.Fraction(9/8)]

```
def rIntervalVector(self,lseq=None)
```
- inter-onset duration interval content of the sequence
  - lseq (list of fractions) – reference list of duration for evaluating interval content – the default list is the same as above.

```
def displayRhythm(self,xml=False,prime=False)
```

Network functions:

```
def rhythmDictionary(Nc,a=None,REF='e')
```
- Generate the dictionary of all possible rhythmic sequences of Nc length in a generalized meter space of N durations
  - Nc (int)– cell length
  - a (str) - list of durations in the rhythm sequence

- returns the dictionary as pandas DataFrame and indicates all non retrogradable cells

```
def rhythmPDictionary(N,Nc,REF='e')
```
- Generate the dictionary of all possible rhythmic sequences from all possible groupings of N REF durations
    - N (int) – number of REF units
    - Nc (int) - cardinality of the grouping
    - returns the dictionary as pandas DataFrame and indicates all non retrogradable cells

```
def rhythmNetwork(input_csv,thup=1.5,thdw=0.0,distance='euclidean',prob=1,w=False):
```
- generate the network of rhythmic cells based on distances between duration vectors
    - input_csv (str) – file containing the dictionary generated by rhythmNetwork
    - thup, thdw (float) – upper and lower thresholds for edge creation
    - distance (str) – choice of norm in the musical space, default is 'euclidean'
    - prob (float)– if ≠ 1, defines the probability of acceptance of any given edge
    - in output it writes the nodes.csv and edges.csv as separate files in csv format

```
def vectorDistance(a,b,distance='euclidean')
```
- calculates the minimal duration distance between two rhythmic cells of same cardinality (bijective)
    - a,b (str) – rhythmic cells

```
def rLeadNetwork(input_csv,thup=1.5,thdw=0.1,w=True,distance='euclidean',prob=1)
```
- generation of the network of all minimal rhythm leadings in a generalized musical space of Nc-dim rhythmic cells – based on the rhythm distance operator
    - input_csv (str) – file containing the dictionary generated by rhythmNetwork
    - thup, thdw (float) – upper and lower thresholds for edge creation
    - w (logical) – if True it writes the nodes.csv and edges.csv files in csv format
    - returns nodes and edges tables as pandas DataFrames

The most computationally intensive parts of the modules can be run on parallel processors using the MPI (Message Passing Interface) protocol. Communications are handled by two additional modules: `communications` and `load_balancing`. Since the user will never have to interact with these modules, we omit here a detailed description of their functions.

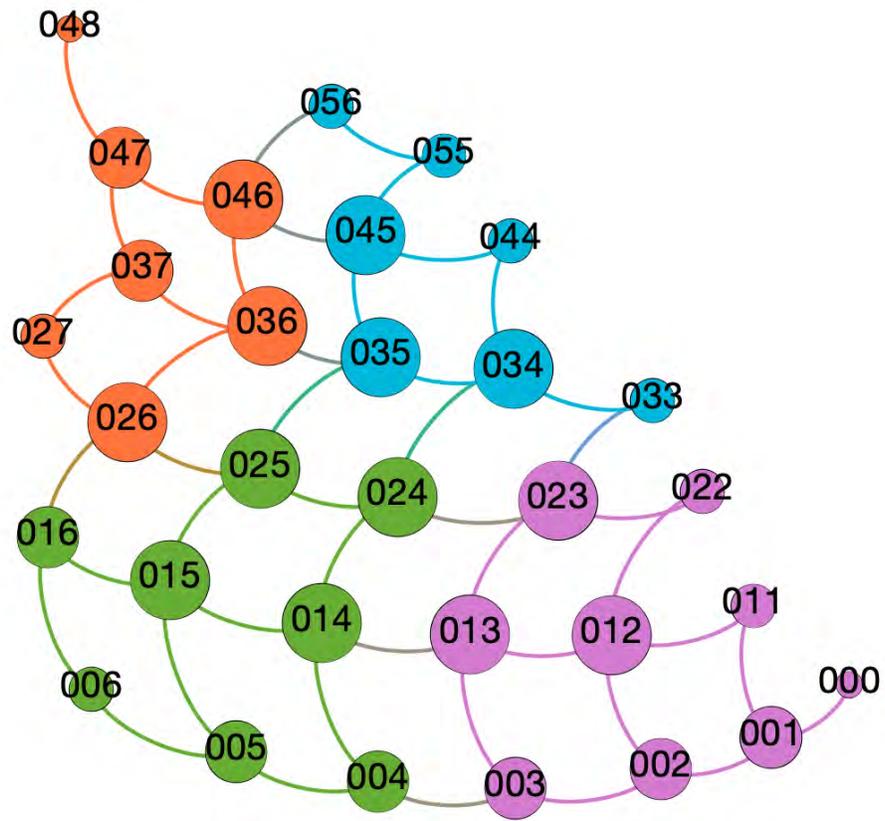

**Fig. S3.**
Nearest neighbor's octave equivalent normal-form triads network in 12TET. The network is restricted to nearest neighbors ($\mathbf{O}(1)$) and is isomorphic to the orbifold of the quotient space $\mathbb{T}^2/\mathcal{S}_3$ as derived in Ref. [5]

**Movie S1.mp4**

Animation of the progressions in the voice leading networks of J.S. Bach's chorale from the Cantata "Erfreut euch, ihr Herzen", BWV 66 and of the sixth movement (sehr langsam) of A. Schoenberg "Sechs kleine Klavierstücke", Op. 19.

**Data S1.gexf (separate file)**

network data file for Figure 1 (main text) in gexf (Gephi) [6] format

**Data S2. gexf (separate file)**

network data file for Figure 2a (main text) in gexf (Gephi) format

**Data S3. gexf (separate file)**

network data file for Figure 2b (main text) in gexf (Gephi) format

**Data S4. gexf (separate file)**

network data file for Figure 2c (main text) in gexf (Gephi) format

**Data S5. gexf (separate file)**

network data file for Figure 2d (main text) in gexf (Gephi) format

**Data S6. gexf (separate file)**

network data file for Figure S2 (SI) in gexf (Gephi) format

**Data S7. gexf (separate file)**

network data file for Figure 3a (main text) in gexf (Gephi) format

**Data S8. gexf (separate file)**

network data file for Figure 4 (main text) in gexf (Gephi) format